\documentclass[12pt]{article}

\usepackage{epsfig,color,graphics,amsmath,multicol,amssymb,rotate}
\usepackage{graphicx}

\begin{document}

\def\rr{ {\vec r} }
\def\RR{ {\bf R} }
\def\jj{ {\bf j} }
\def\ee{ {\bf e} }
\def\FF{ {\bf F} }
\def\qq{ {\bf q} }
\def\kk{ {\vec k} }
\def\FF{ {\bf F} }
\def\EE{ {\bf E} }
\def\VV{ {\bf V} }
\def\UU{ {\bf U} }
\def\RR{ {\bf R} }
\def\vv{ {\bf v} }
\def\uu{ {\bf u} }
\def\ttt{ {\bf t} }
\def\hh{ {\bf h} }
\def\XX{ {\bf X} }
\def\xx{ {\bf x} }
\def\mh{ {\underline{h}}}
\def\meta{{\underline{\epsilon}} }
\def\mt{ {\underline{\tau}}\  }
\def\mht{ {\underline{h}^T} }
\def\meps{ {\underline{\epsilon}}\ }
\def\msig{ {\underline{\sigma}}\  }
\def\mSigma{ {\underline{\Sigma}}\  }
\def\mG{ {\underline{G}} }
\def\m1{ {\underline{1}} }
\def\mg{ {\underline{g}} }
\def\rhohat{ {\hat{\rho}} }
\def\br{{\bf r}}

\begin{center}
 {\bf  INTRODUCING VARIABLE CELL SHAPE METHODS IN FIELD THEORY SIMULATIONS OF POLYMERS}

\vspace{1cm}

         Jean-Louis BARRAT$^{a,b}$, Glenn H. FREDRICKSON$^{b}$, Scott W.
         SIDES$^{b}$

 \vspace{1cm}

 $^{a}$     \textit{Laboratoire de Physique de la Mati\`ere Condens\'ee et Nanostructures,
            Universit\'e Claude Bernard Lyon I and CNRS, 6 rue Amp\`ere, 69622 Villeurbanne Cedex, France}

$^{b}$ \textit{Department of Chemical Engineering \& Materials
Research Laboratory, University of California, Santa Barbara, CA
93106, USA}

\end{center}

\vspace{1cm}

{\bf ABSTRACT} We propose a new method for carrying out
field-theoretic simulations of polymer systems under conditions of
prescribed external stress, allowing for shape changes in the
simulation box. A compact expression for the deviatoric stress
tensor is derived in terms of the chain propagator, and is used to
monitor changes in the box shape according to a simple relaxation
scheme. The method allows fully relaxed, stress free
configurations to be obtained even in non trivial morphologies,
and enables the study of morphology transitions induced by
external stresses.

 \centerline{\bf last modified \today}

\section{Introduction}
In particle-based simulations, such as Molecular Dynamics (MD) or
Monte Carlo (MC), it has long been realized that the flexibility
of the approach could be greatly improved by imposing intensive,
rather than extensive, thermodynamic constraints. Schemes to
impose pressure are well known in Monte Carlo simulations, and
constant temperature and constant pressure algorithms, using
various thermostats and barostats,  have become common options in
Molecular Dynamics simulations \cite{Frenkel}. When dealing with
solids, however, it was pointed out by Parrinello and Rahman
\cite{Parrinello81} that the study of structural transformations
would necessitate a release of the constraints imposed on the
symmetry of the simulation cell. They introduced, within the
context of MD simulations, a method that treated the cell shape as
a dynamical variable, coupled to an externally imposed stress
tensor. The method was later slightly modified by Ray and Rahman
\cite{Ray84} to account for some subtleties associated with
describing large deformations. The Parrinello-Rahman-Ray technique
has been used in a variety of contexts, e.g. to describe
structural transformations between crystalline phases in solids
and to compute elastic constants under pressure, and is still
evolving as new applications are identified
\cite{Martins97,Parrinello03}.

In the polymer literature, such variable cell shape methods have
been applied in particle based simulations (see e.g.
\cite{Brown93}). However, in the other prominent class of
simulation methods for complex fluids, i.e. the
``field-theoretic'' approaches \cite{ghf02}, no equivalent of the
variable shape simulation cell methods has been proposed up to
now. The most familiar type of field-theoretic simulation is a
numerical implementation of self consistent field theory (SCFT)
\cite{schmid98,matsen94}, which amounts to the computation of a
saddle point (mean-field) configuration of a statistical field
theory model. Modern SCFT simulations are normally of two types.
The so-called \emph{large cell} calculations are used to address
bulk fluid systems with no long-range order, or microphase
separated polymeric fluids where the symmetry of the periodic
domain structure is not known in advance. Cubic simulation cells
with periodic boundary conditions are normally applied for this
purpose \cite{Sides03,drolet99,fraaije97} and large cells must be
used to avoid finite-size effects. Often the cell size is adjusted
in such calculations to minimize the the free energy, although the
cell shape is fixed. In the second type of SCFT simulation, a
so-called \emph{unit cell} calculation, one assumes a particular
crystallographic symmetry and applies a simulation cell that
corresponds to a primitive cell compatible with that symmetry
\cite{matsen94,Matsen}. In these calculations, the cell size
\emph{and} shape are adjusted to minimize the free energy of a
unit cell of the periodic structure \cite{tyler03b}. No analogous
procedure currently exists for adjusting the cell shape in large
cell calculations in order to achieve stress-free conditions.

In view of the growing interest in the computation of the
mechanical properties of the complex mesophases that can be formed
in copolymer melts \cite{Tyler03,Thompson04}, it is of interest to
formulate a method similar to the Parrinello-Ray-Rahman approach
within the framework of field-based polymer simulations. Such a
``variable cell shape'' approach should provide a useful option in
SCFT calculations when dealing with systems subjected to
anisotropic stresses (e.g. unaxial extension or shear), or in
situations where the optimal symmetry of the simulation box is not
\textit{a priori} obvious. Even more simply, it would provide a
simple means of relaxing the box size and shape to an optimal
value, simply by imposing a 'zero stress' external condition. In
the present manuscript, we present a first step towards such a
simulation methodology in which the cell shape is adjusted at
fixed cell volume. Such a formalism is particularly appropriate
for field theory models that invoke the incompressible liquid
assumption.

The paper is organized as follows. In Section \ref{general}, we
recall some useful  notations and results from elasticity theory.
The expression for the stress tensor appropriate for polymeric
systems is discussed in Section \ref{stress_sec}, where we also
describe the equations used for evolving the shape of the
simulation cell. Some examples of two-dimensional simulations for
AB diblock copolymer melts are given in Section \ref{examples}.

\section{General Notations}
\label{general}

In this section, we closely follow the description made by Ray and
Rahman \cite{Ray84} of the Parrinello-Rahman type methods used in
particle simulations. In a system with periodic boundary
conditions, the simulation cell can be defined by three (in
general nonorthogonal) independent  vectors
$\hh_1$,$\hh_2$,$\hh_3$ forming the sides of the parallelepiped
cell. The Cartesian coordinates of these vectors can be used to
construct  a $3\times3$ matrix $\mh$ defined by $\mh = (\hh_1 ,
\hh_2 , \hh_3 )$. The Cartesian coordinates of any point $\RR$ in
the cell can be expressed as
\begin{equation} \label{rx}
\RR = \mh \XX
\end{equation}
 where $\XX$ is a rescaled vector whose components
lie in $[0,1]$. Integrals on $\RR$ can be converted into integrals
over $\XX$ by using a scaling factor $\det \mh$, which represents
the volume of the cell, $V$. In the case of a particle or monomer
number density, for example, one can write
\begin{equation}
\rho(\RR)= \rho(\XX) ( \det \mh)^{-1}
\end{equation}

The metric tensor $\mG$ is constructed from $\mh$ as
\begin{equation}
\mG = \mht \mh
\end{equation}
where $\mht$ is the transpose of $\mh$.  $\mG$ is used in
transforming dot products from the original Cartesian to rescaled
coordinates, according to
\begin{equation}
\RR\cdot \RR^\prime = \XX \cdot \mG \cdot \XX^\prime = X_\alpha
G_{\alpha \beta} X^\prime_\beta
\end{equation}
where here and in the following summation over repeated indexes is
implicit.

Elasticity theory describes the deformation of any configuration
from a reference configuration in terms of a strain tensor. This
tensor is constructed by relating the vector connecting two points
in the deformed configuration to the corresponding displacement of
the same points in the reference configuration. If the reference
configuration of the simulation box is denoted by $\mh_0$,  the
strain is given by
\begin{equation}\label{strain}
\meta = \frac{1}{2}\left[ (\mht_0)^{-1} \mht  \mh (\mh_0)^{-1}
-\m1 \right] = \frac{1}{2}\left[ (\mht_0)^{-1} \mG (\mh_0)^{-1}
-\m1 \right]
\end{equation}
where $\m1$ denotes the unit tensor. It is important to note that
this expression, usually known as the ``Lagrangian stress tensor''
is not limited to small deformations \cite{landau}. Usually, the
reference configuration $\mh_0$ will be defined as a state of the
system under zero applied external stress.

The thermodynamic variable conjugate to this strain tensor, in the
sense that the elementary  work done on the system can be written
in the form
\begin{equation}\label{dw}
\delta W = V_0 \mathrm{Tr} (\mt \meta),
\end{equation}
is the thermodynamic \emph{tension} tensor $\mt$ \cite{wallace70},
also known as Piola-Kirchhoff second stress tensor. $V_0 \equiv
\det \mh_0$ denotes the volume of the system in the reference
configuration. This thermodynamic tension tensor can be related to
the more usual \emph{Cauchy stress tensor} $\msig$ through
\begin{equation}\label{tsigma}
    \msig = \frac{V_0}{V}   \mh \: (\mh_0 )^{-1} \mt (\mht_0 )^{-1}\mht
\end{equation}
The tension is the derivative of the free energy with respect to
the strain, which is calculated from the reference configuration.
The stress, on the other hand, is the derivative of the free
energy with respect to an incremental strain taken with respect to
the actual configuration. The difference between these two
quantities can be understood, qualitatively, from the fact that
the strain is not an additive quantity, as can be seen from the
existence of the nonlinear term in eq \ref{strain}. While the
Cauchy stress has a mechanical meaning in terms of forces applied
to the sample, the thermodynamic tension is a purely thermodynamic
quantity, and does not in general have a simple mechanical
interpretation.

It was proposed by Ray and Rahman \cite{Ray84} that, in a particle
system, an appropriate ensemble to carry out simulations allowing
for shape transformations in the simulation cell is a constant
tension, constant temperature, constant number of particles
ensemble, usually denoted as a ``$\tau$TN'' ensemble. In this
ensemble, fluctuations in the shape and size of the simulation
cell, described by $\mh$, are allowed, and the configurational
partition function is written as
\begin{equation}\label{z1}
    Z= \int d\mh \int d^N\RR \exp(-\beta V_0 \mt:\meta - \beta H(\RR))
\end{equation}
where $\RR$ denotes the coordinates of the $N$  particles, which
evolve in the ``box'' defined by $\mh$, $\beta \equiv 1/(k_B T)$,
and $H(\RR)$ is the potential energy. In eq \ref{z1}, no kinetic
energy terms in the Hamiltonian associated with either particle or
cell motion were included. Such terms are integral in the
Parrinello-Rahman method in order to obtain equations of motion
for the  cell and the particles. In SCFT theory, however, the
inertial particle dynamics is replaced by overdamped relaxational
dynamics for the density fields, and we will make use of a similar
relaxation scheme for the cell coordinates. Thus, the kinetic
energy terms prove to be irrelevant.

\section{Application to Polymer Field Theory}
\label{stress_sec}

\subsection{Polymer Partition Function}
The key step in formulating a polymer field theory is to transform
the integral over the particle coordinates appearing in the
partition function for a complex fluid (c.f. equation \ref{z1})
into an integral over density and/or chemical potential fields for
the different species. This step has been described in many papers
(see e.g. \cite{ghf02,matsen94}), so we will only highlight here
the technical modifications associated with a variable cell shape
simulation method. To illustrate the method, it is convenient to
work with a standard mesoscopic model of an incompressible AB
diblock copolymer melt \cite{matsen94}. In this model, the melt
consists of $n$ identical diblock copolymer chains composed of
monomer species $A$ and $B$ and contained in a volume $V$. Each of
the chains has a total of $N$ statistical segments; a fraction $f$
of these segments are type $A$ and constitute the A block of each
macromolecule. For simplicity, the volume occupied by each
segment, $v_0$, and the statistical segment length, $b$, are
assumed to be the same for the A and B type segments. The
Hamiltonian for this system can be written
\begin{equation}\label{h0}
H= \sum_{i=1}^{n} \frac{k_BT}{4 R_{g0}^2} \int_0^1 ds
\left(\frac{d\RR_i(s)}{ds}\right)^2 + v_0 \chi_{AB} k_BT \int d\br
\: \rhohat_A(\br) \rhohat_B(\br)
\end{equation}
where $\RR_i (s)$ with $s \in [0,1]$ is a space curve describing
the conformation of the $i$th copolymer and $R_{g0}^2 =b^2 N/6$ is
the radius of gyration of an ideal chain of $N$ statistical
segments. Interactions between dissimilar segments A and B are
described by the Flory parameter $\chi_{AB}$. The densities
$\rhohat_{A,B}(\br)$ are microscopic segment density fields
defined by
\begin{equation}\label{eq1}
    \rhohat_A (\br ) = N \sum_{i=1}^n \int_0^f ds \; \delta (\br -
    \RR_i (s) )
\end{equation}
and
\begin{equation}\label{eq2}
    \rhohat_B (\br ) = N \sum_{i=1}^n \int_f^1 ds \; \delta (\br -
    \RR_i (s) )
\end{equation}
A local incompressibility constraint $\rhohat_A (\br ) + \rhohat_B
(\br ) = \rho_0$ is imposed in this standard copolymer melt model
for all points $\br$ in the simulation domain. The total segment
density $\rho_0$ can evidently be expressed as $\rho_0 =nN/V =
1/v_0$.

In order to implement the Parrinello-Ray-Rahman method, it is
convenient to switch from the Cartesian coordinates $\RR$ to the
rescaled coordinates $\XX$, as defined in eq \ref{rx}. Using the
rescaled coordinates, the Hamiltonian of eq \ref{h0} is expressed
as
\begin{eqnarray}
\beta H & = & \sum_{i=1}^n \frac{1}{4 R_{g0}^2} \int_0^1 ds
\frac{dX_{i\alpha}(s)}{ds}  G_{\alpha\beta}
\frac{dX_{i\beta}(s)}{ds} \nonumber \\
& + & v_0 \chi_{AB} (\det\mh)^{-1} \int d\xx \: \rhohat_A(\xx)
\rhohat_B(\xx)
\end{eqnarray}
where the coordinates $\XX$ and $\xx$ are taken in $[0,1]^3$. The
partition function is
\begin{eqnarray}
 Z & = & \int d(\mh)  (\det \mh)^{nN} \delta (\det \mh - V_0 ) \:  \exp(-\beta V_0 \mt:\meta)  \nonumber \\
 & \times & \prod_{i=1}^n \int \mathcal{D}X_i \exp(-\beta H) \prod_{\xx} \delta ( \rhohat_A (\xx ) +\rhohat_B (\xx
 ) -nN) \label{partf}
\end{eqnarray}
The final factor in the above expression imposes the constraint of
local incompressibility for the scaled density fields. Moreover,
the assumption of incompressibility implies globally that the cell
volume remains fixed at its initial value, i.e. $\det \mh = V =
V_0 = \det \mh_0$. This is enforced by the delta function in the
first line above. All shape transformations considered in the
present work are therefore volume preserving. The practical
implementation of this constraint will be discussed below.

Hubbard-Stratonovich transformations are used to convert the
particle-based partition function \ref{partf} into a statistical
field theory \cite{ghf02}. These can be carried out
straightforwardly on the polymer partition function for a given
cell shape $\mh$, $Z(\mh )$, with the result
\begin{eqnarray}
Z(\mh) & \equiv  &  \prod_{i=1}^n \int \mathcal{D}X_i \exp(-\beta
H) \prod_{\xx} \delta ( \rhohat_A (\xx ) +\rhohat_B (\xx
 ) -nN) \nonumber \\
 & = & \int \mathcal{D}w \; \exp(  n \ln Q[w,\mh]-E[w]) \label{zpol}
 \end{eqnarray}
 where $Q[w,\mh]$ is the partition function of a single copolymer chain experiencing a
 chemical potential field $w (\xx ,s )$, $\int \mathcal{D}w$ denotes a functional
 integral over the field $w$, and $E[w]$ is a local quadratic functional of $w$ that reflects
 the A-B monomer interactions and the local incompressibility constraint: \cite{ghf02}
\begin{equation}\label{eqghf1}
    E[w] = \frac{n}{2} \int d\xx \; \left[ \frac{1}{2 \chi N} (w_B
    -w_A )^2 - (w_A + w_B ) \right]
\end{equation}
Here we have noted that for an AB diblock copolymer melt, the
potential $w(\xx ,s)$ amounts to a two-component potential, i.e.
$w(\xx ,s) =w_A (\xx )$ for $s \in [0,f]$ and $w(\xx ,s) =w_B (\xx
)$ for $s \in [f,1]$.

The object $Q[w,\mh]$ is a normalized partition function for a
single copolymer experiencing a potential field $w(\xx ,s)$ and is
defined as
\begin{equation}\label{eqghf2}
    Q[w,\mh ] \equiv \frac{\int \mathcal{D} \XX \: e^{-\frac{1}{4
    R_{g0}^2} \int_0^1 ds \: \frac{d \XX}{ds} \cdot \mG \cdot \frac{d
    \XX}{ds} - \int_0^1 ds \: w(\XX (s) ,s)}}{\int \mathcal{D} \XX \: e^{-\frac{1}{4
    R_{g0}^2} \int_0^1 ds \: \frac{d \XX}{ds} \cdot \mG \cdot \frac{d
    \XX}{ds}}}
\end{equation}
The path integral in the denominator ensures the normalization
$Q[0,\mh]=1$, but it depends only on $\det \mh =V$, so it remains
constant for volume-preserving changes in cell shape $\mh$. The
partition function $Q$ can be obtained from a single-chain
propagator $q (\xx ,s)$ that is the solution of a modified
diffusion equation
\begin{equation}
\frac{\partial q}{\partial s} = R_{g0}^2 (\mG^{-1})_{\alpha\beta}
\frac{\partial^2 q}{\partial x_\alpha \partial x_\beta} -w(\xx,s)
q(\xx,s) \label{ghfeq11}
\end{equation}
subject to $q(\xx ,0)=1$. The single chain partition function is
given by $Q[w,\mh ]= \int d\xx \; q(\xx,1)$.

By virtue of the above equations, the partition function for an
incompressible diblock copolymer melt confined to a cell of
variable shape can be expressed as a field theory in the variables
$\mh$ and $w$:
\begin{equation}
 Z  =  \int d(\mh) \int \mathcal{D}w \; \delta (\det \mh - V_0 ) \:
  \exp (-F[w,\mh ])
  \label{ghfeq3}
\end{equation}
where $F[w,\mh]$ is an effective Hamiltonian given by
\begin{equation}\label{ghfeq4}
    F[w,\mh] = \beta V_0 \mt:\meta +E[w] - n \ln Q[w,\mh ]
\end{equation}
The factor $(\det \mh )^{nN} =V_0^{nN}$, which was present in eq
\ref{partf}, has no thermodynamic significance for a constant
volume system, so it has been absorbed into the integration
measure in eq \ref{ghfeq3}. This equation provides a rather
general starting point for carrying out variable cell shape
simulations of polymer field theories. Here we shall invoke the
mean-field approximation (SCFT) and, for a given shape $\mh$ of
the simulation box, approximate the functional integral over $w$
in eq \ref{ghfeq3} by the saddle point method. For this purpose,
the functional $Q[w,\mh ]$ can be evaluated for any $w$ and $\mh$
by solving the modified diffusion equation using a parallel
implementation of a pseudospectral operator splitting scheme
\cite{rasmu02,Sides03}. The saddle point (mean-field) value of
$w$, $w^*$, is obtained by applying a relaxation algorithm
\cite{ghf02,Sides03,ghfsws03} to solve
\begin{equation}\label{ghfeq9}
    \left. \frac{\delta F[w,\mh ]}{\delta w(\xx ,s)}
    \right|_{w=w^*} =0
\end{equation}
In the mean-field approximation, $F[w^* ,\mh ]$ corresponds to the
free energy of the copolymer melt (in units of $k_B T$).

\subsection{Equation of Motion for the Cell}

In the Parrinello-Rahman method, the cell is ascribed a fictitious
kinetic energy, and the corresponding Hamilton equations are
solved along with the equations of motion for the particles. As
the equations of motion for the fields in SCFT are overdamped
relaxation equations, a reasonable choice is to use a similar
relaxational dynamics to evolve the cell coordinates. We have
implemented the following dynamics for the evolution of $\mh$:
\def\uD{\underline{D}}
\def\uM{\underline{M}}
\def\uI{\underline{1}}
\begin{equation}\label{evolveh}
\frac{d \mh}{dt} = -\lambda_0 \: \mh  \uD \mh^{-1} \frac{\partial
F[w,\mh ]}{\partial \mh}
\end{equation}
where the tensor $\uD$ is a projection operator whose action on an
arbitrary tensor $\uM$ is a traceless tensor, i.e. $\uD \: \uM
\equiv \uM - (1/3) \mathrm{Tr} (\uM ) \uI$. Equation \ref{evolveh}
corresponds to a cell shape relaxation that (for $\lambda_0 >0$)
is down the gradient $\partial F /\partial \mh$, approaching a
local minimum of the mean-field free energy $F[w^* ,\mh ]$. The
``mobility'' tensor $\mh \uD \mh^{-1}$ is chosen so that the cell
shape dynamics described by eq \ref{evolveh} conserves the cell
volume. This can be seen from the identity
\begin{equation}\label{ghfeq5}
    \frac{d}{dt} \ln \det \mh = \mathrm{Tr} (\mh^{-1} \frac{d}{dt}
    \mh )
\end{equation}
which implies that $\det \mh$ is a constant of the motion if
$\mh^{-1} (d/dt) \mh$ is traceless.

Application of eq \ref{evolveh} requires an expression for the
thermodynamic force $\partial F/\partial \mh$. Explicit
differentiation, noting the constraint of constant $\det \mh$,
leads to
\begin{equation}\label{ghfeq6}
    \frac{\partial F[w,\mh ]}{\partial \mh} = \beta V_0
    \left( \frac{\partial}{\partial \mh}\mathrm{Tr} (\meta \mt)
 + \mh \underline{\Sigma} \right)
\end{equation}
where $\underline{\Sigma}$ is a symmetric tensor defined by
\begin{eqnarray}
   \Sigma_{\alpha\beta} [w,\mh ] &  = & - \frac{2 k_B T n}{V}
   \frac{\partial \ln Q[w,\mh ]}{\partial G_{\alpha \beta}}
   \nonumber \\
   & = & \frac{k_BT n}{2V R_{g0}^2}
   \left< \int_0^1 ds
   \frac{dX_{\alpha} (s)}{ds}\frac{dX_{\beta} (s)}{ds}\right> \label{stress}
\end{eqnarray}
The angular brackets in the second expression denote an average
over all conformations $\XX (s)$ of a single copolymer chain that
is subject to a prescribed chemical potential field $w$ and fixed
cell shape $\mh$.

The first term on the right hand side of eq \ref{ghfeq6} can be
conveniently rewritten as
\begin{equation}
\frac{\partial}{\partial h_{\alpha\beta} }(\mathrm{Tr} (\meta
\mt)) = \frac{\partial}{\partial h_{\alpha\beta} }
(\mathrm{Tr}{1\over 2} \mht_0^{-1} \mht\mh \: \mh_0^{-1} \mt) =
(\mh \: \mh_0^{-1}\mt \mht_0^{-1})_{\alpha\beta}.
\end{equation}
Hence, eq \ref{evolveh} can be compactly expressed
as\begin{equation}\label{evolveh1}
    \frac{d\mh}{dt} = -\lambda \mh \uD \left[(\mh_0^{-1}\mt \mht_0^{-1})
    + \mSigma\right]
\end{equation}
where $\lambda >0$ is a new relaxation parameter defined by
$\lambda = \beta V_0 \lambda_0$.

Equation \ref{evolveh1} is a purely relaxational equation that
will evolve the cell shape to a configuration of minimum free
energy (in the mean-field approximation). This configuration can
either be metastable (local minimum) or stable (global minimum).
Addition of a noise source to the equation provides a means for
overcoming free energy barriers between metastable and stable
states, i.e. a simple simulated annealing procedure. Beyond
mean-field theory, in order to fully assess the effects of thermal
fluctuations in both $w$ and $\mh$, a more sophisticated procedure
is required that takes into account the non-positive definite
nature of the polymer field theory \cite{ghf02}.

An equilibrium solution of the cell shape equation \ref{evolveh1}
is evidently obtained when
\begin{equation}\label{equil}
(\mh_0^{-1}\mt \mht_0^{-1}) +  \mSigma = 0
\end{equation}
Combining eqs \ref{tsigma}, \ref{stress} and \ref{equil}, it is
easily seen that this equilibrium condition corresponds to a
balance between the \emph{externally} applied Cauchy stress,
$\msig$, and the \emph{internal} elastic stress, $\msig^{int}$,
sustained by the polymer chains
\begin{equation}\label{ghfeq10}
    \msig + \msig^{int} =0
\end{equation}
where
\begin{equation}\label{smallsig}
\sigma_{\alpha\beta}^{int} [w,\mh] \equiv (\mh \mSigma \mht
)_{\alpha\beta} = \frac{k_B T}{2V R_{g0}^2} \sum_{i=1}^{n}
   \left< \int_0^1 ds \frac{dR_{i\alpha}}{ds}\frac{dR_{i\beta}}{ds} \right> \ .
\end{equation}
This expression for the internal polymer stress is well-known in
the polymer literature \cite{doiedwards}.

Equation \ref{evolveh1} drives a change in the shape of the
simulation cell (at constant cell volume) to approach the
equilibrium condition \ref{equil} at which the internal elastic
stress of the copolymers balances the imposed external stress. In
practice, our numerical strategy consists in implementing a cell
shape evolution at regular intervals, typically every 50 to 100
field minimization steps. The magnitude of the coefficient
$\lambda$ in eq \ref{evolveh1} should ensure that the strain
associated with such evolutions remains moderate. As the driving
force $(\mh_0^{-1}\mt \mht_0^{-1}) + \mSigma$ is normally of order
unity (in the appropriate reduced units used in SCFT calculations,
i.e. taking $k_BT$ as the unit of energy and $R_{g0}$ as the unit
of length), $1/\lambda$ is a number of the order of the interval
between cell shape evolutions.

A final technical point is that the change in cell shape should
ideally be performed at a fixed spatial resolution for the
solution of the diffusion equation. Hence, a change in the cell
shape generally involves re-meshing the rescaled simulation cell
with a grid that will be finer in the expanded directions and
coarser in the compressed directions, in order to keep the mesh
size locally invariant.

\subsection{Expression for the Stress Tensor}

The last step in the implementation is to find an expression for
the internal stress tensor $\msig^{int}$ (eq \ref{smallsig}) or
$\mSigma$ (eq \ref{stress}) in terms of the single chain
propagator, which is the central object computed in a
field-theoretic simulation \cite{ghf02,Fredrickson02}. In order to
obtain such an expression, we return to the definition of
$\mSigma$
\begin{equation}\label{stress1}
 \beta V \Sigma_{\alpha\beta} = - 2 n
  \frac{1}{Q}\frac{\partial Q[w,G]}{\partial G_{\alpha \beta}}
\end{equation} The derivative of the single chain
partition function can be calculated by discretizing the paths
with a small contour step $\Delta$.
\begin{eqnarray}\label{}
 -\frac{2}{Q}\frac{\partial Q[w,G]}{\partial G_{\alpha \beta}}=&
 \frac{1}{2R_{g0}^2 Q} \int_0^1 ds \int d\XX \int d\XX'\cr
& \int \mathcal{D}\XX(s) \delta(\XX-\XX(s))
 \delta(\XX'-\XX(s+\Delta))
 \left(\frac{X_\alpha-X_\alpha'}{\Delta}\right)
 \left(\frac{X_\beta-X_\beta'}{\Delta}\right)\cr
&\exp\left(-\frac{1}{4 R_{g0}^2}\int_0^1 ds
\frac{dX_{\alpha}(s)}{ds}  G_{\alpha\beta}
\frac{dX_{\beta}(s)}{ds} - \int_0^1 ds \: w(\XX(s),s)\right)
 \end{eqnarray}
Except between the points $\XX,s$ and $\XX',s+\Delta$ one can
replace the path integrals with propagators $q$, so that
\begin{eqnarray}\label{}
 -\frac{2}{Q}\frac{\partial Q[w,G]}{\partial G_{\alpha \beta}}=&
\frac{1}{2R_{g0}^2 Q} \int_0^1 ds \int d\XX \int d\XX' q(\XX,s)
q(\XX',1-s-\Delta) \cr
 &\left(\frac{X_\alpha-X_\alpha'}{\Delta}\right)
 \left(\frac{X_\beta-X_\beta'}{\Delta}\right)
 \exp\left( -\frac{1}{4\Delta R_{g0}^2}G_{\alpha\beta}(X_\alpha-X_\alpha')(X_\beta-X_\beta')\right) \cr
 \end{eqnarray}

One can then set $\XX'=\XX+\uu$, and expand for small $\uu$ and
$\Delta$ according to
\begin{equation}\label{}
q(\XX+\uu,1-s-\Delta) = q(\XX,1-s) -\Delta \frac{\partial
q}{\partial s} + u_\gamma \frac{\partial q}{\partial X_\gamma }+
\frac{1}{2}u_\gamma u_\delta \frac{\partial^2 q}{\partial X_\gamma
\partial X_\delta}
\end{equation}
The derivative w.r.t. $s$ can be eliminated by applying the
modified diffusion eq \ref{ghfeq11}.  One also requires second and
fourth moments of the Gaussian distribution of displacements
$\uu$,
 $$\overline{u_\alpha u_\beta} = G^{-1}_{\alpha\beta} (2 R_{g0}^2 \Delta)$$
 $$\overline{u_\alpha u_\beta u_\gamma u_\delta}=
 (2 R_{g0}^2 \Delta)^2 (G^{-1}_{\alpha\beta}G^{-1}_{\gamma\delta}
+G^{-1}_{\alpha\gamma}G^{-1}_{\beta\delta} +
G^{-1}_{\alpha\delta}G^{-1}_{\beta\gamma})$$

By means of these results, we have
\begin{eqnarray}\label{}
 -\frac{2}{Q}\frac{\partial Q[w,G]}{\partial G_{\alpha \beta}}=&
 G^{-1}_{\alpha\beta}\frac{ \int d\XX\rho(\XX)}{\Delta}
 +  G^{-1}_{\alpha\beta} \int d\XX w(\XX) \rho(\XX) \cr
 - &\frac{R_{g0}^2}{Q} G^{-1}_{\alpha\beta} G^{-1}_{\gamma\delta} \int d\XX \int_0^1 ds
 q(\XX,s)\frac{\partial^2 q}{\partial x_\gamma \delta
 x_\beta}\cr
 + &
 \frac{R_{g0}^2}{Q} (G^{-1}_{\alpha\beta}G^{-1}_{\gamma\delta}+
G^{-1}_{\alpha\gamma}G^{-1}_{\beta\delta} +
G^{-1}_{\alpha\delta}G^{-1}_{\beta\gamma})\int d\XX \int_0^1 ds
 q(\XX,s) \frac{\partial^2 q}{\partial x_\gamma \partial
 x_\delta}\cr &
\end{eqnarray}
where $\rho(\XX) =Q^{-1} \int_0^1 ds \: q(\XX ,s)q(\XX,1-s)$ is
the single-chain total monomer density operator. There is a
partial cancellation in the last two terms so that
\begin{eqnarray}\label{}
 -\frac{2}{Q}\frac{\partial Q[w,G]}{\partial G_{\alpha \beta}}=&
 G^{-1}_{\alpha\beta}\frac{ \int d\XX\rho(\XX)}{\Delta}
 + i G^{-1}_{\alpha\beta} \int d\XX w(\XX) \rho(\XX) \cr
 + &
 \frac{2 R_{g0}^2}{Q}
G^{-1}_{\alpha\gamma}G^{-1}_{\beta\delta} \int d\XX \int_0^1 ds
 q(\XX,s) \frac{\partial^2 q}{\partial X_\gamma \partial
 X_\delta}\cr &
\end{eqnarray}

The internal polymer stress is obtained after matrix
multiplication by $\mh$ on the left and $\mht$ on the right. This
implies that the first two terms become a simple isotropic stress
contribution, and are therefore not relevant to an incompressible
system. The final formula for the internal stress tensor is
therefore, apart from this diagonal contribution,
\begin{equation}\label{stressfinal}
    \sigma_{\alpha \beta}^{int} =\left(\frac{nk_BT}{V}\right)\frac{2 R_{g0}^2}{Q}
h^{-1}_{\gamma \alpha} \int d\XX \int_0^1 ds \;
 q(\XX,s) \frac{\partial^2 q(\XX,1-s)}{\partial X_\gamma \partial
 X_\delta} h^{-1}_{\delta\beta}
\end{equation}
The tensor $\Sigma$ appearing in equation  \ref{evolveh1} is given
by an
 expression similar to \ref{stressfinal}, with $\mG$ replacing $\mh$.
 The factor $k_B T n/V$ accounts for the total number of chains,
and produces a stress with the correct dimensions.  In practice,
the stress will be made dimensionless by dividing by this factor,
so that the dimensionless stress is given by
\begin{equation}\label{stressfinal1}
   \frac{ \sigma_{\alpha \beta}^{int}}{(n/V) k_B T} =
   \frac{2 R_{g0}^2}{ Q}
h^{-1}_{\gamma \alpha} \int d\XX \int_0^1 ds \;
 q(\XX,s) \frac{\partial^2 q(\XX,1-s)}{\partial X_\gamma \partial
 X_\delta} h^{-1}_{\delta\beta}
\end{equation}
An equivalent expression that better reveals the symmetry of the
stress tensor is obtained by an integration by parts:
\begin{equation}\label{stressfinal2}
   \frac{ \sigma_{\alpha \beta}^{int}}{(n/V) k_B T} = -
   \frac{2 R_{g0}^2}{ Q}
h^{-1}_{\gamma \alpha} \int d\XX \int_0^1 ds \;
 \frac{\partial q(\XX,s)}{\partial X_\gamma} \frac{\partial q(\XX,1-s)}{\partial
 X_\delta} h^{-1}_{\delta\beta}
\end{equation}
A local (rather than volume averaged) version of this connection
between the stress tensor and the polymer propagator was derived
previously in \cite{Fredrickson02}. Numerically, $\sigma_{\alpha
\beta}$ is evaluated from eq \ref{stressfinal1} using the
pseudo-spectral scheme of \cite{rasmu02,ghfsws03}. The derivatives
with respect to spatial coordinates are obtained by multiplying
the propagator by the appropriate components of the wavevector in
Fourier space, and transforming back into real space.

\section{Examples}
\label{examples}

\subsection{Stress-Strain Relation for Lamellae}

\begin{figure}
\begin{center}
\includegraphics[height=8cm]{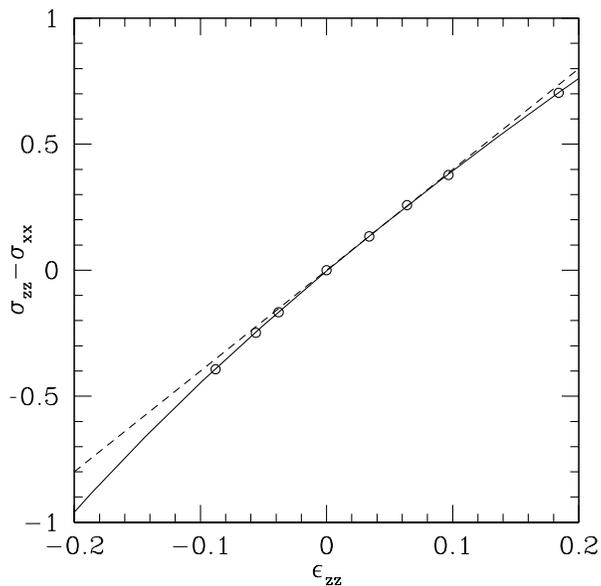}
\end{center}
\caption{Stress-strain relation for perfectly aligned lamellae,
where the stress is expressed in units of $(n/V) k_B T$ according
to eq \ref{stressfinal1}. The full line is obtained from imposed
strain simulations, while the dots are obtained using constant
tension simulations. The dashed line corresponds to the linear
elastic behavior around the zero stress configuration. The
associated elastic constant is $B=4(\rho/N)k_BT$ } \label{fig1}
\end{figure}

As a first example and check of the method, we consider the
stress-strain relations for a perfectly aligned lamellar system,
with lamellae perpendicular to the $z$ direction. Such a system
can support a stress only in the $z$ direction and is therefore
described by a single elastic relation, $\sigma_{zz}$ versus
$\epsilon_{zz}$ (Other  ``apparent'' elastic constants computed in
\cite{Thompson04} can be shown to vanish, as is obvious from eq
\ref{stressfinal} in a system that is translationally invariant
along the $x$ and $y$ directions). For such a system, we have
computed the stress-strain relation using either an imposed strain
method or an imposed stress, as described above. The imposed
strain amounts to progressively deforming the simulation cell
along the $z$ direction, keeping the topology (i.e. the number of
lamellae) fixed. The stress is then evaluated using eq
\ref{stressfinal}. The reference configuration was created by
running the variable shape simulation code under zero stress
conditions (zero tension), in order to obtain a perfectly relaxed
initial configuration. The results, shown in Figure \ref{fig1},
indicate that the two methods produce identical results in this
simple situation. Finally, we should mention that lamella under
dilational stress should give rise to a ``chevron'' instability at
rather small strains, as described in \cite{wang93}. In order to
avoid the development of this instability, the equations were
solved without any added noise, which allows one to preserve the
metastable lamellar topology even at large strains.

\subsection{Formation of Cylindrical Phase}

\begin{figure}
\begin{center}
\begin{tabular}{ccc}
\hspace{-5cm} \includegraphics[height=12cm,angle=-90]{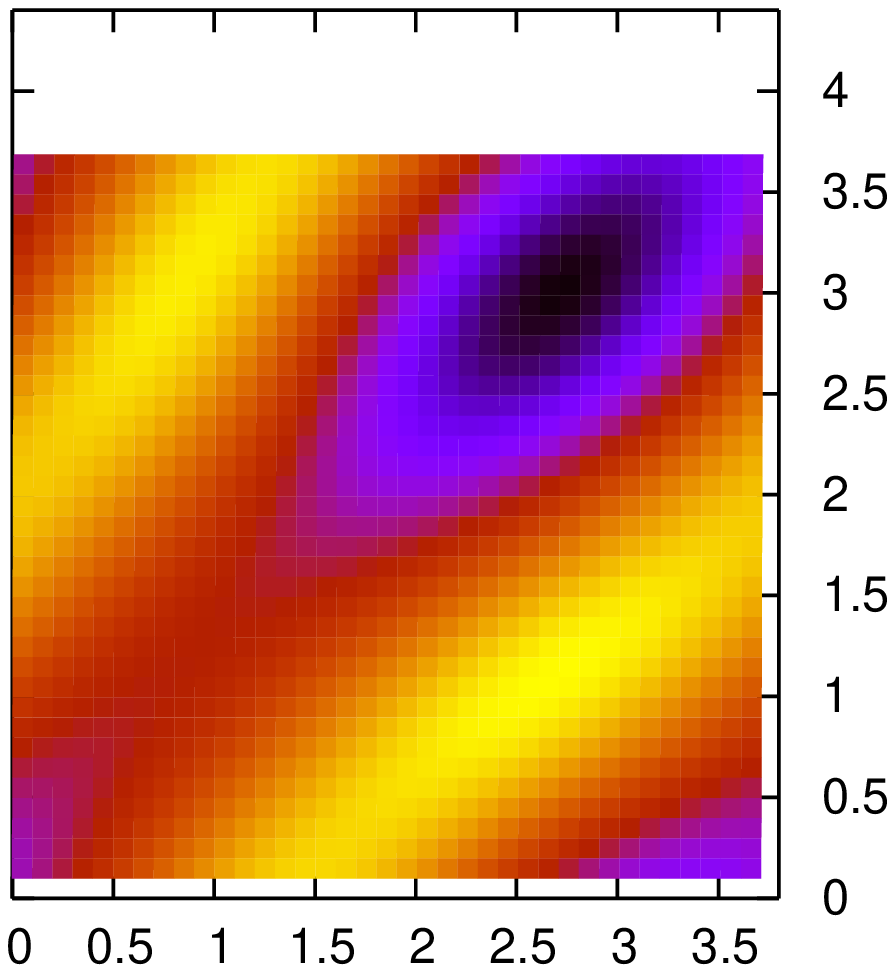}&
\hspace{-8cm}  \includegraphics[height=12cm,angle=-90]{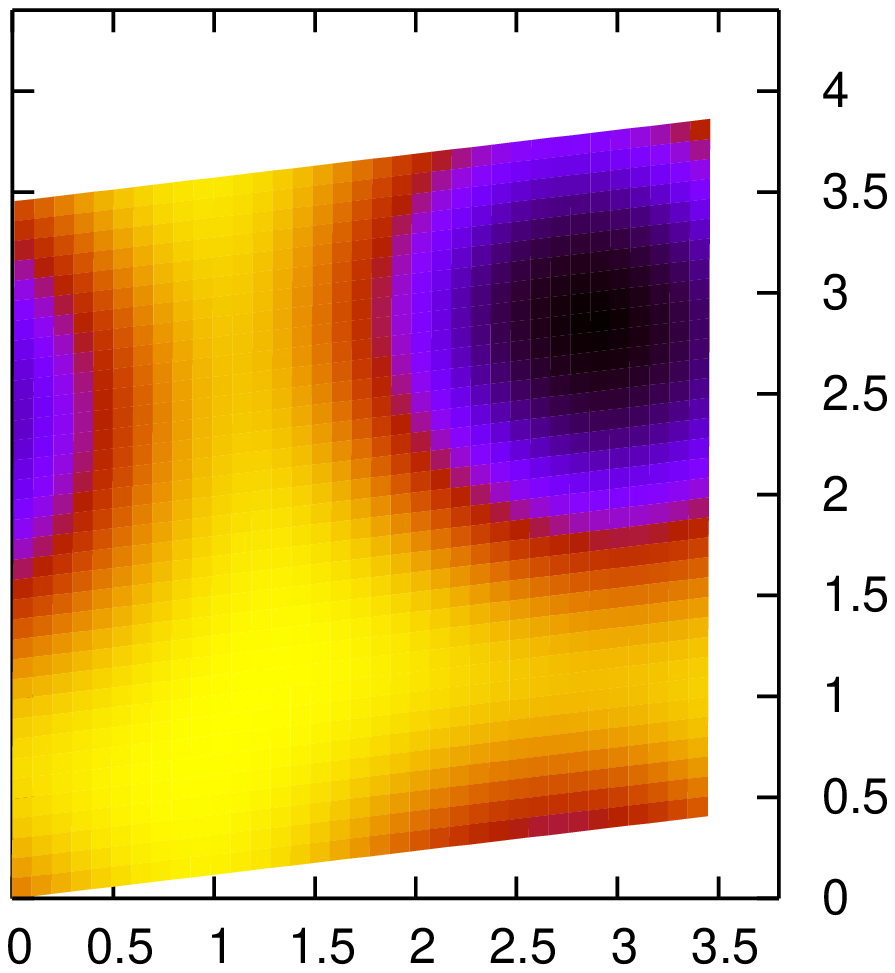}&
\hspace{-8cm} \includegraphics[height=12cm,angle=-90]{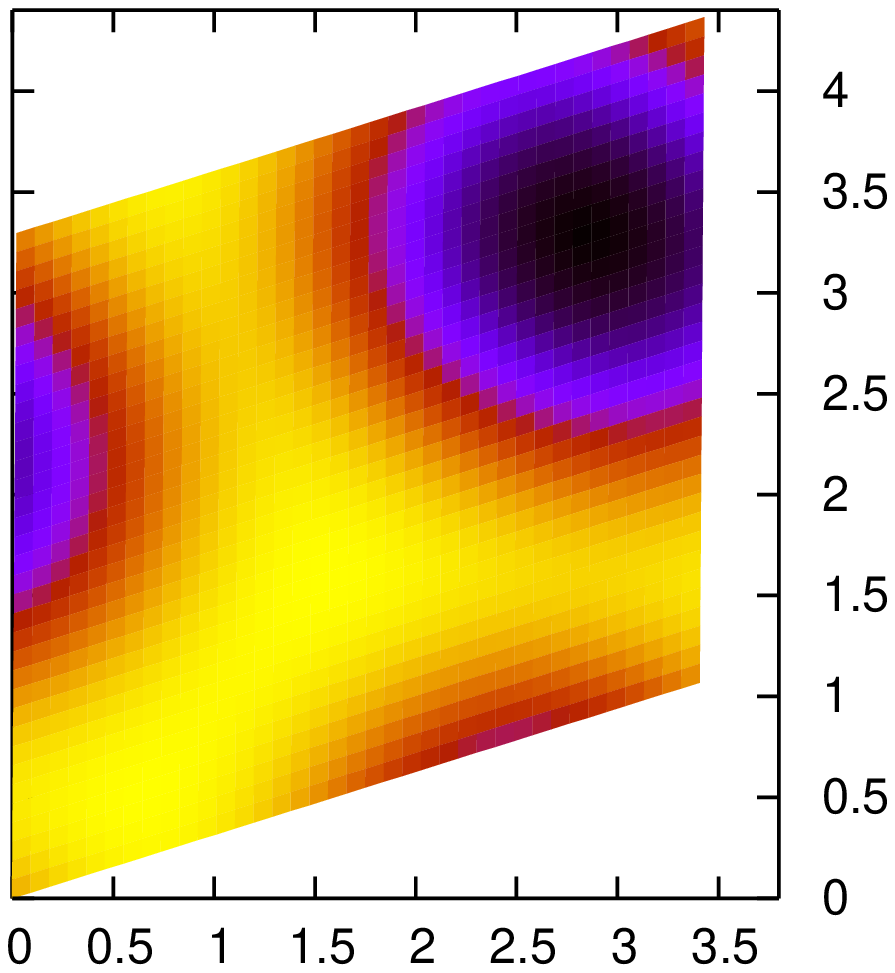}
\\
\end{tabular}
\end{center}
\caption{Transformation of a square cell under zero external
stress, when the melt is quenched into the stability region of the
cylindrical phase ($\chi N=15.9$, $f=0.64$).} \label{fig2}
\end{figure}

As a second example, we show how a square cell, quenched into the
region of the phase diagram where the equilibrium phase is a
hexagonally-packed array of cylinders, naturally evolves under
zero tension to a rhombic shape characteristic of a primitive cell
of the Bravais lattice. Starting from a random initial field
configuration and imposing a quench to $\chi N=15.9$ (for a
diblock copolymer with $f=0.64$), successive stages of this
transformation are shown in Figure \ref{fig2}. A cylindrical
nucleus appears after a few iterations, and the cell progressively
deforms from a square to a rhombus consistent with the triangular
organisation of the lattice. Relaxation to this equilibrium shape
proceeds in typically 1000 iterations, with a change in cell shape
implemented every 10 iterations.

\subsection{Shift of the Lamellae-Cylinder Transition Under Stress}

\begin{figure}
\begin{center}
\includegraphics[height=8cm]{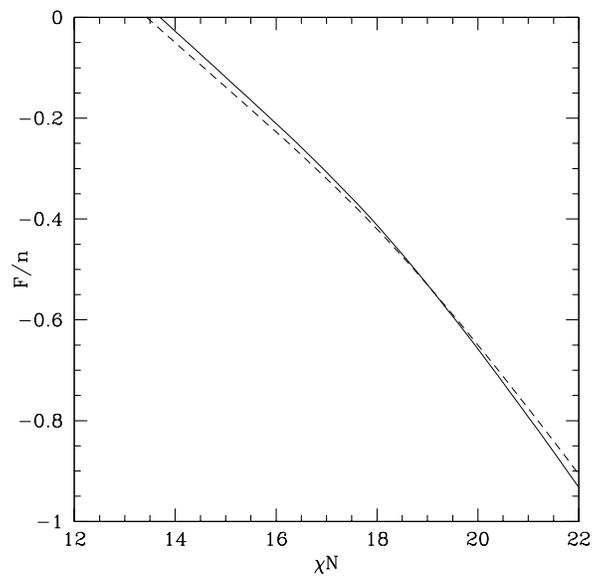}
\end{center}
\caption{Free energies (per chain) of the lamellar (solid) and
cylindrical (dashed) phases under zero tension, as a function of
$\chi N$, for a diblock with $f=0.64$.} \label{fig3}
\end{figure}

As a final application example, we attempt to address the
following question: how does a compressive stress affect the
lamellar to hexagonal transition in a two dimensional system? We
first determine the location of this transition for a simple $AB$
diblock copolymer melt with the fraction $f$ of $A$ monomers equal
to $f=0.64$ by running a ``zero stress'' simulation for various
values of $\chi N$. As the lamellar spacing or triangular lattice
spacing change with $\chi N$, the zero stress simulation
represents a convenient way of automatically adjusting the cell
dimensions to the minimum free energy, thus obtaining fully
relaxed, stress free configurations at each value of $\chi N$. The
resulting transition takes place at $\chi N\simeq 18$, as shown in
Figure \ref{fig3}, which was obtained by ramping $\chi N$ at zero
tension.

\begin{figure}
\begin{center}
\begin{tabular}{cccccc}
\hspace{-6cm} \includegraphics[height=12cm,angle=-90]{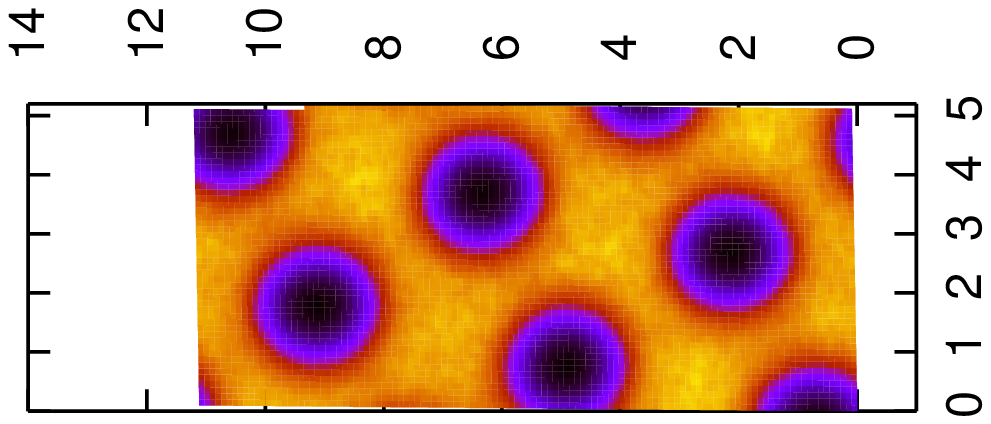}&
\hspace{-10cm} \includegraphics[height=12cm,angle=-90]{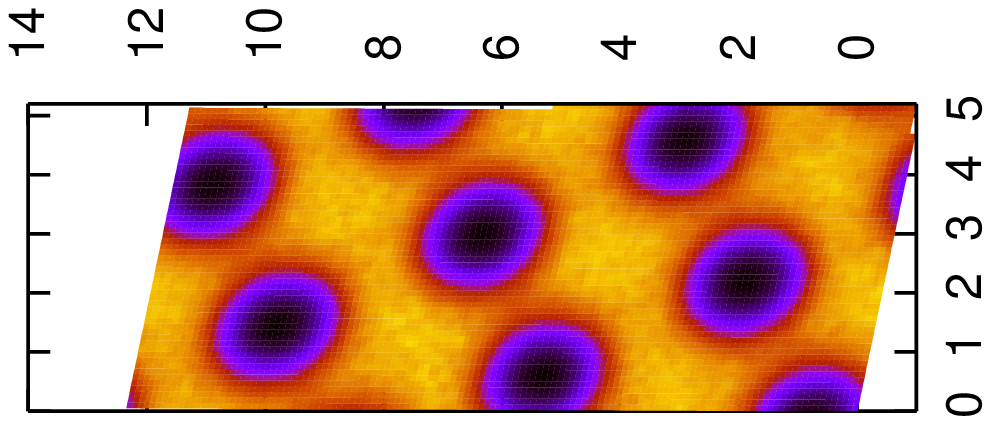}&
\hspace{-10cm} \includegraphics[height=12cm,angle=-90]{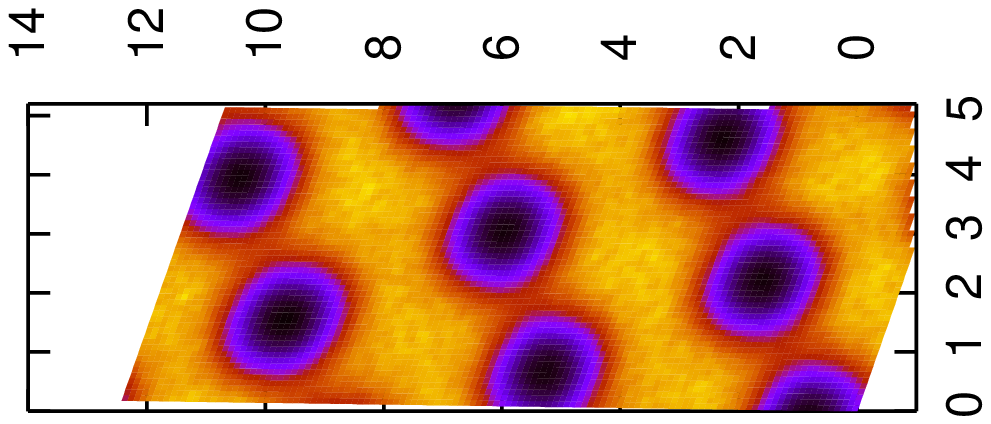}&
\hspace{-10cm} \includegraphics[height=12cm,angle=-90]{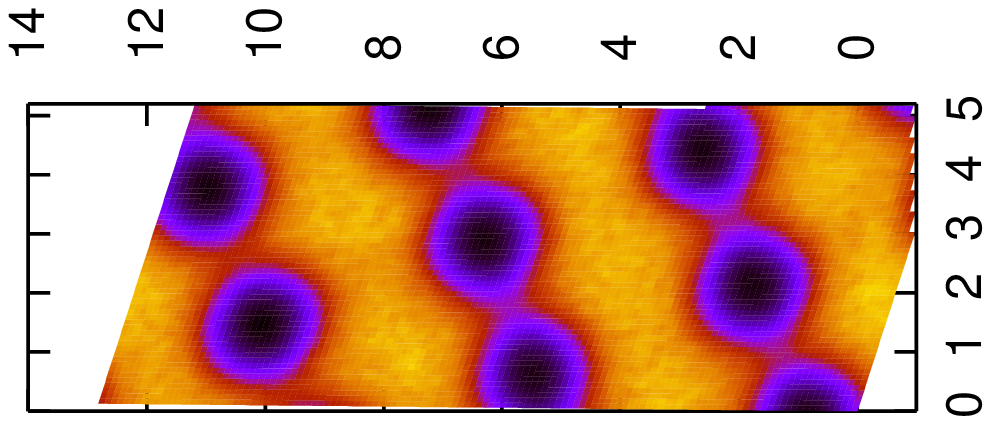}&
\hspace{-10cm} \includegraphics[height=12cm,angle=-90]{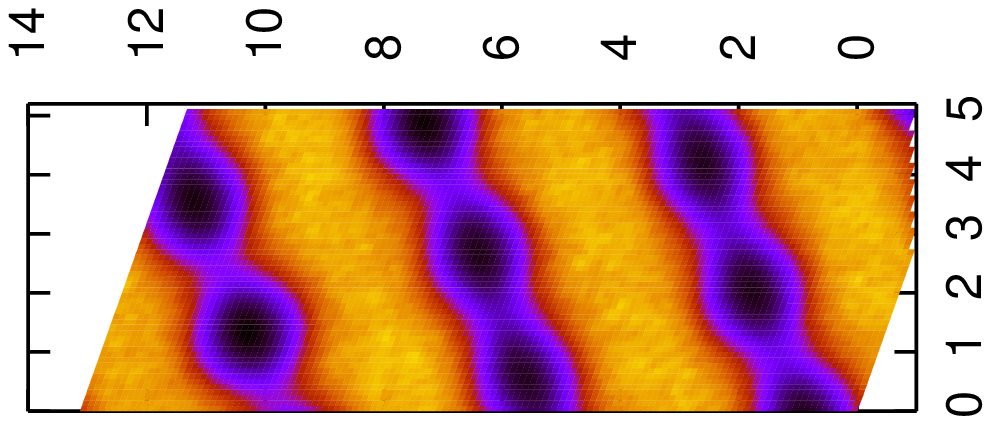}&
\hspace{-10cm} \includegraphics[height=12cm,angle=-90]{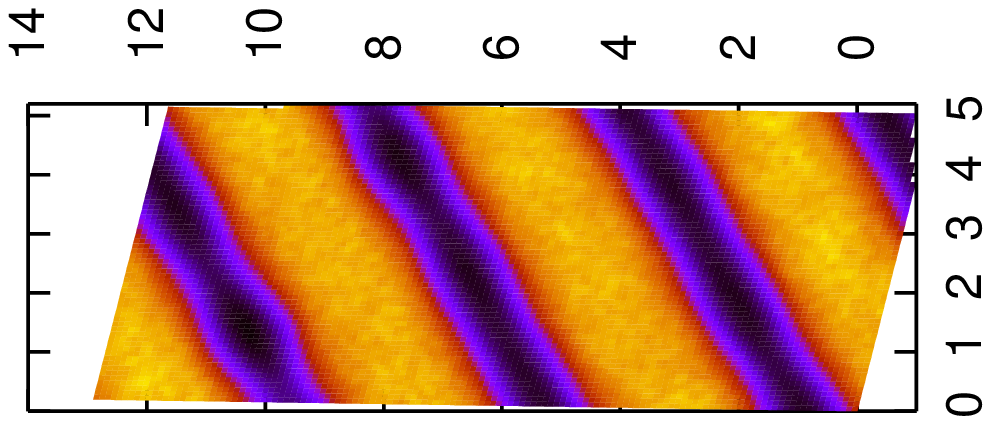}\\
\end{tabular}
\end{center}
\caption{Six stages of the cylindrical-lamellar transformation
upon compression of a cylindrical phase at $\chi N= 16$ along the
horizontal axis ($\sigma_{xx}-\sigma_{yy}=1$).} \label{fig4}
\end{figure}

As the lamellar spacing at the transition is smaller than the
spacing between cylinders of the triangular lattice, it can be
expected that the application of a compressive stress will favor
the lamellar phase over the hexagonal one. In order to verify this
hypothesis, we have carried out simulations in which the hexagonal
phase at $\chi N=16$ is subjected to an external uniaxial
compressive stress of $1$ in reduced units. Under this imposed
compression, the initially almost rectangular cell shape
transforms into a parallelogram, until suddenly the structural
transformation takes place (see Figure \ref{fig4}). Comparing the
free energies under tension, it is found that the lamellar phase
has indeed a lower free energy when the transition takes place.
However, since the resulting lamella are not aligned with the
principal directions of the stress, they cannot sustain the
resulting shear so that the cell eventually elongates without
bound.

\section{Conclusion and perspectives}

Variable shape simulations have played, and are still playing, an
important role in molecular simulations of solids for the
identification of allotropic phase transformations under pressure
and of the associated  transformation pathways. Here we have shown
that the methods used in solids can be generalized to simulations
of polymer field theories. In this context, many types of
polymeric fluids can be treated as incompressible fluid mixtures.
As a first step, we have therefore focused our attention on the
incompressible case and developed a methodology by which a
simulation cell can be adjusted in shape to attain the lowest free
energy in the mean-field (SCFT) approximation. In the case of zero
applied stress, the method provides a very convenient way of
obtaining mechanical equilibrium in large systems, which may
include defects or complex structures. We have also shown that the
application of a non-hydrostatic tension can modify the
thermodynamic equilibrium among phases, and that a variable shape
simulation cell allows the system to explore non-trivial pathways
for structural transformations.

The relaxational sampling of cell shapes that we have used in this
paper is rudimentary, and there is certainly room for improvement
if one is interested in studying, e.g., nucleation barriers
between phases. Another interesting perspective is that, although
we have been treating the internal stress tensor as a global
property of the system, it is clear from eq \ref{stressfinal} (see
also \cite{Fredrickson02}) that a local value of the stress tensor
can be easily identified prior to volume integration. This formula
should be useful in assessing the local stress distribution in
inhomogeneous systems under strain, including polymer-based
composites.

{\bf Acknowledgments} Useful discussions with Drs. S. Bauerle and
K. Katsov are gratefully acknowledged. J-L. Barrat thanks the
Materials Research Laboratory for supporting his stay at UCSB, and
section 15 of CNRS for the grant of a temporary research position
during this visit. This work was supported by the MRSEC Program of
the National Science Foundation under Award No. DMR00-80034.

%%%%%%%%%%%%%%%%%%%%%%%%
\bibliographystyle{unsrt}

\begin{thebibliography}{10}

\bibitem{Frenkel}
D. Frenkel and B. Smit,   \textit{Understanding Molecular
Simulation} (Academic, New York, 1996)

\bibitem{Parrinello81}  M. Parrinello and A. Rahman, J. Appl. Phys. \textbf{52},
7182 (1981)

\bibitem{Ray84} J. R. Ray and A. Rahman, J. Chem. Phys. \textbf{80},
4423 (1984)

\bibitem{Klein} T.H.K. Barron, M.L. Klein, Proc. Phys. Soc.
{\bf 65}, 523 (1965)

\bibitem{Martins97} I. Souza, J.L. Martins, Phys. Rev B \textbf{55}, 8733
(1997)

\bibitem{Parrinello03} R. Martonák, A. Laio, M. Parrinello, Phys. Rev. Lett. \textbf{90}, 075503 (2003)

\bibitem{Brown93}
 J. I. McKechnie, R. N. Haward, D. Brown, J. H. R. Clarke,
 Macromolecules, \textbf{26}, 198
(1993)

\bibitem{ghf02} G.H. Fredrickson, V. Ganesan, and F. Drolet,
Macromolecules, \textbf{35}, 16 (2002)

\bibitem{schmid98} F. Schmid, J. Phys.: Condens. Matter
\textbf{10}, 8105 (1998)

\bibitem{matsen94} M.W. Matsen, M. Schick, Phys. Rev. Lett,
\textbf{72}, 2660 (1994)

\bibitem{Sides03} S.W. Sides, G.H. Fredrickson, Polymer, \textbf{44}, 5859
(2003)

\bibitem{drolet99} F. Drolet, G.H. Fredrickson, Phys. Rev. Lett,
\textbf{83}, 4317 (1999)

\bibitem{fraaije97} N.M. Maurits, J.G.E.M. Fraaije, J. Chem.
Phys., \textbf{107}, 5879 (1997)

\bibitem{Matsen} M.W. Matsen, J. Chem. Phys, \textbf{106}, 7781 (1997)

\bibitem{tyler03b} C.A. Tyler, D.C. Morse, Macromolecules \textbf{36}, 8184
(2003).
\bibitem{Tyler03} C.A. Tyler, D.C. Morse, Macromolecules \textbf{36}, 3764 (2003).

\bibitem{Thompson04} R. B. Thompson, K. Ø. Rasmussen, T. Lookman
J. Chem. Phys. \textbf{101}, 3990 (2004)

\bibitem{landau} L. Landau, E.M. Lifschitz, \textit{Theory of
Elasticity} (Pergamon Press, New-York, 1986)


\bibitem{wallace70} D. Wallace, \textit{Thermodynamics of crystals }(Wiley,
New-York 1973)

\bibitem{rasmu02} K.O. Rasmussen and G. Kalosakas, J. Polym. Sci.,
Part B: Polym. Phys. \textbf{40}, 1777 (2002).

\bibitem{ghfsws03} G.H. Fredrickson and S.W. Sides, Macromolecules
\textbf{36}, 5415 (2003).

\bibitem{doiedwards} M. Doi, S.F. Edwards, \textit{The Theory of Polymer
Dynamics} (Oxford University Press, Oxford, 1988)

\bibitem{Fredrickson02} G.H. Fredrickson, J. Chem. Phys, \textbf{117}, 6810 (2002)

\bibitem{wang93} Z.G. Wang, J. Chem. Phys. \textbf{100}, 2298 (1993)



\end{thebibliography}

\end{document}